\documentclass[twocolumn,showpacs,superscriptaddress,amsmath,amssymb,aps,pra]{revtex4-2}
\usepackage{graphicx}
\usepackage{CJKutf8}
\usepackage{dcolumn}
\usepackage{amsmath,bm}
\usepackage{epstopdf}
\usepackage[mathlines]{lineno}
\usepackage{color}
\usepackage[colorlinks=true,linkcolor=blue,citecolor=magenta,urlcolor=cyan]{hyperref}

\usepackage{tikz,xcolor,hyperref}

\definecolor{lime}{HTML}{A6CE39}
\DeclareRobustCommand{\orcidicon}{%
	\begin{tikzpicture}
	\draw[lime, fill=lime] (0,0) 
	circle [radius=0.16] 
	node[white] {{\fontfamily{qag}\selectfont \tiny ID}};
	\draw[white, fill=white] (-0.068,0.105) 
	circle [radius=0.007];
	\end{tikzpicture}
	\hspace{-2mm}
}

\foreach \x in {A, ..., Z}{%
	\expandafter\xdef\csname orcid\x\endcsname{\noexpand\href{https://orcid.org/\csname orcidauthor\x\endcsname}{\noexpand\orcidicon}}
}

\begin{document}
\title{Searching strong `spin'-orbit coupled one-dimensional hole gas in strong magnetic fields}
\author{Rui\! Li~(\begin{CJK}{UTF8}{gbsn}李睿\end{CJK})\orcidA{}}
\email{ruili@ysu.edu.cn}
\affiliation{Key Laboratory for Microstructural Material Physics of Hebei Province, School of Science, Yanshan University, Qinhuangdao 066004, China}
\date{\today}

\begin{abstract}
We show that a strong `spin'-orbit coupled one-dimensional (1D) hole gas is achievable via applying a strong magnetic field to the original two-fold degenerate (spin degeneracy) hole gas confined in a cylindrical Ge nanowire. Both strong longitudinal and strong transverse magnetic fields are feasible to achieve this goal. Based on quasi-degenerate perturbation calculations, we show the induced low-energy subband dispersion of the hole gas can be written as $E=\hbar^{2}k^{2}_{z}/(2m^{*}_{h})+\alpha\sigma^{z}k_{z}+g^{*}_{h}\mu_{B}B\sigma^{x}/2$, a form exactly the same as that of the electron gas in the conduction band. Here the Pauli matrices $\sigma^{z,x}$ represent a pseudo spin (or `spin' ), because the real spin degree of freedom has been split off from the subband dispersions by the strong magnetic field. Also, for a moderate nanowire radius $R=10$ nm, the induced effective hole mass $m^{*}_{h}$ ($0.065\sim0.08~m_{e}$) and the `spin'-orbit coupling $\alpha$ ($0.35\sim0.8$ eV~\AA) have a small magnetic field dependence in the studied magnetic field interval $1<B<15$ T, while the effective $g$-factor $g^{*}_{h}$ of the hole `spin' only has a small magnetic field dependence in the large field region.
\end{abstract}
\maketitle

\section{Introduction}
There are well developed techniques for initialization, manipulation, and readout of the electron spin states in gate-defined semiconductor quantum dots~\cite{RevModPhys.79.1217,vandersypen2019}, such that the quantum dot electron spin has been regarded as one of the most promising qubit candidates for implementing quantum computations~\cite{loss1998quantum}. Owing to a suppressed interaction between the hole spin and the lattice nuclear spins, quantum dot hole spin is also expected to be an excellent qubit candidate as well as the electron spin~\cite{Scappucci:2020aa}. Meanwhile, the band dispersions near the top of the valence band of semiconductors are described by the Luttinger-Khon Hamiltonian~\cite{PhysRev.97.869,PhysRev.102.1030}, where there is a large intrinsic spin-orbit coupling~\cite{PhysRevLett.98.097202,Wang:2021wc}, such that the quantum dot hole spin has the advantage of being manipulable by an external oscillating electric field.

Planar (2D)~\cite{Hendrickx:2020ab,Hendrickx:2020aa} or nanowire (1D)~\cite{PhysRevLett.101.186802,PhysRevLett.112.216806,Froning:2021aa,Gao2020AM,Zhang:2021wj} hole quantum dot can be fabricated experimentally via placing proper metallic gates below a 2D or 1D hole gas. Note that the physics of the hole spin qubit in a planar quantum dot may be totally different from that in a nanowire quantum dot. Take the recent extensively studied semiconductor Ge as an illustration~\cite{PhysRevB.88.241405,Watzinger:2018aa,PhysRevB.94.041411,Watzinger:2016aa,PhysRevB.102.205412,PhysRevResearch.3.013081}, the lowest subband dispersion of the 2D hole gas in a Ge quantum well always has heavy hole character~\cite{PhysRevB.103.125201}, and can be modeled by a parabolic curve with band minimum at the center of the $k$ space~\cite{winkler2003spin}. While the low-energy subband dispersions of the 1D hole gas in a cylindrical Ge nanowire are quite different~\cite{PhysRevB.84.195314}. The lowest two subband dispersions of the 1D hole gas anticross with each other at the center of the $k_{z}$ space~\cite{PhysRevB.84.195314,RL_2021}, and the shape of these two dispersions is very similar to that of a strong spin-orbit coupled 1D electron gas described by the Hamiltonian $H_{c}=p^{2}/(2m^{*}_{e})+\alpha\sigma^{z}p+g_{e}\mu_{B}B\sigma^{x}/2$~\cite{lirui2018energy,lirui2018the}. However, there is an additional spin degeneracy in the hole subband dispersions~\cite{RL_2021}. We note that the above strong spin-orbit coupled electron gas model has many applications in the studies of the spin-orbit qubits~\cite{trif2008spin,lirui2013controlling,PhysRevB.87.205436,PhysRevB.92.054422,PhysRevApplied.14.014090,lirui2018a,PhysRevB.99.014308,Li2020charge}, the Bose-Einstein condensations~\cite{PhysRevLett.108.225301,PhysRevLett.110.235302,PhysRevA.91.023604}, the Kondo physics of a spin-orbit coupled quantum wire~\cite{PhysRevB.94.125115,Lopes_2020,LOPES2019188,PhysRevB.102.155114}, and the Majorana fermions~\cite{PhysRevLett.105.077001,PhysRevLett.105.177002,PhysRevB.90.195421}. 

In this paper, we are inspired to achieve a strong spin-orbit coupled 1D hole gas in a cylindrical Ge nanowire, which would share similar potential applications with the electron gas. In order to achieve a pure (without the hole spin degeneracy) `spin'-orbit coupled 1D hole gas, we apply a strong magnetic field to split off the unwanted spin degree of freedom from the hole subband dispersions. Note that the `spin' here is more properly regarded as a pseudo spin, it is introduced to describe the induced low-energy subband dispersion of the hole gas in a strong magnetic field in comparison with the conduction band electron case. Because both the longitudinal and the transverse $g$-factors of the hole gas at the band minimum are finite and are comparable to each other~\cite{RL_2021}, such that both strong longitudinal and strong transverse magnetic fields are feasible to achieve this splitting goal. The induced low-energy hole subband dispersion is completely the same as that of the electron gas in the conduction band, i.e., described by $E=\hbar^{2}k^{2}_{z}/(2m^{*}_{h})+\alpha\sigma^{z}k_{z}+g^{*}_{h}\mu_{B}B\sigma^{x}/2$. A large `spin'-orbit coupling $\alpha$ ($\sim0.8$ eV~\AA) of the Rashba type is achievable for a moderate nanowire radius $R=10$ nm. Note that stronger Rashba spin-orbit coupling ($\sim2$ eV~\AA) for electrons have been reported in Pb-atomic wires~\cite{Brand:2015wn}  and Te-atomic chains~\cite{han2020giant}. The magnetic field dependences of the hole effective mass $m^{*}_{h}$, the strength of `spin'-orbit coupling $\alpha$, and the effective $g$-factor $g^{*}_{h}$ (for the pseudo hole spin) are discussed in details.

\section{1D hole gas}
Here we are interested in a 1D hole gas confined in a cylindrical Ge nanowire of radius $R$ in the absence of the external magnetic field. The axis direction of the nanowire is defined as the $z$-direction. For semiconductor Ge in the bulk, the band dispersions near the top of the valence band are well described by the Luttinger-Kohn Hamiltonian in the spherical approximation~\cite{PhysRev.102.1030,WU201061}. Hence, using the language of the effective mass approximation, we write the Hamiltonian of a hole confined in this nanowire as~\cite{PhysRevB.84.195314,PhysRevB.79.155323,PhysRevB.78.033307}
\begin{equation}
H_{0}=\frac{1}{2m_{e}}\left[\left(\gamma_{1}+\frac{5}{2}\gamma_{s}\right)\textbf{p}^{2}-2\gamma_{s}(\textbf{p}\cdot\textbf{J})^{2}\right]+V(r),\label{Eq_model}
\end{equation}
where $m_{e}$ is the bare electron mass, $\gamma_{1}=13.35$ and $\gamma_{s}=(2\gamma_{2}+3\gamma_{3})/5=5.11$ are Luttinger parameters~\cite{PhysRevB.4.3460} for semiconductor Ge, ${\bf p}=-i\hbar\nabla$ is the momentum operator, ${\bf J}=(J_{x}, J_{y}, J_{z})$ is a spin-$3/2$ vector operator, and $V(r)$ is the transverse ($xy$ plane) confining potential of the hole
\begin{equation}
V(r)=\left\{\begin{array}{cc}0,~&~r<R,\\
\infty,~&~r>R,\end{array}\right.\label{Eq_potential}
\end{equation}
with $R$ being the radius of the Ge nanowire. It should be noted that in our following calculations, we have chosen a representative and experimentally achievable nanowire radius $R=10$ nm~\cite{PhysRevLett.101.186802,PhysRevLett.112.216806}.

\begin{figure}
\includegraphics{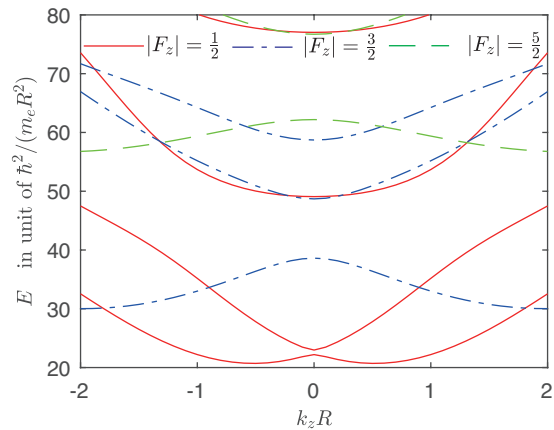}
\caption{\label{fig_subbands}The low-energy subband dispersions of the 1D hole gas. For a cylindrical Ge nanowire with radius $R=10$ nm, the energy unit is $\hbar^{2}/(m_{e}R^{2})\approx0.763$ meV. Note that there is an energy gap ($\approx0.774$) at $k_{z}R=0$ between the first and the second lowest subband dispersions. Also, each line is two-fold degenerate, i.e., spin degeneracy.}
\end{figure}

The model (\ref{Eq_model}) is exactly solvable, the detailed description of the solving method can be found elsewhere~\cite{sweeny1988hole,PhysRevB.42.3690,RL_2021}. The $z$-component of the total angular momentum $F_{z}=-i\partial_{\varphi}+J_{z}$ is a conserved quantity $[F_{z},H_{0}]=0$, such that we can classify the eigenfunctions of $H_{0}$ using $F_{z}$~\cite{PhysRevB.42.3690,RL_2021}. Following the method introduced in Refs.~\cite{sweeny1988hole,PhysRevB.42.3690}, we obtain the low-energy subband dispersions of the 1D hole gas which are explicitly shown in Fig.~\ref{fig_subbands}. As one can see clearly from the figure, the band minimum is not at the center of the $k_{z}$ space. Instead, there are two symmetrical minimums approximately located at $|k_{z}R|\approx0.517$. Also, for wave vectors in the interval $|k_{z}R|<1$, the lowest two subband dispersions, i.e., given by total angular momentum $|F_{z}|=1/2$ (see Fig.~\ref{fig_subbands}), are approximately separated from the other higher subband dispersions.

The shape of the lowest two subband dispersions of the hole gas shown in Fig.~\ref{fig_subbands} is very similar to that of a strong spin-orbit coupled 1D electron gas, e.g., described by the Hamiltonian $H_{c}=p^{2}/(2m^{*}_{e})+\alpha\sigma^{z}p+g_{e}\mu_{B}B\sigma^{x}/2$~\cite{lirui2018energy,lirui2018the}. However, there is an additional spin degeneracy in the hole subband dispersions~\cite{RL_2021}, while there is no degeneracy for the electron case. Note that the spin degeneracy in the hole subband dispersions is a direct consequence of the coexistence of the time-reversal symmetry and the spin-rotation symmetry of the model (\ref{Eq_model}). Here, we are motivated to search the strong `spin'-orbit coupled 1D hole gas via applying a strong magnetic field to lift out the unwanted spin degeneracy in the hole subband dispersions (see Fig.~\ref{fig_subbands}). As we have demonstrated in our previous paper~\cite{RL_2021}, both the longitudinal and the transverse $g$-factors of the hole gas at the band minimum have finite values and are comparable to each other, such that both longitudinal and transverse magnetic fields are feasible to achieve this goal.

For a given energy eigenvalue $E_{n}(k_{z})$ at a given wave vector $k_{z}$, we can obtain the corresponding eigenfunction via fixing the total angular momentum $F_{z}$, e.g., see Figs.~3-5 of Ref.~\cite{RL_2021}. Once one eigenfunction, e.g. $\Psi_{n,k_{z},\Uparrow}$ ($n$ is the subband index), is obtained, the other degenerate counterpart, i.e., $\Psi_{n,k_{z},\Downarrow}$, can be obtained via a combination of the time-reversal and the spin-rotation transformations~\cite{RL_2021}. At a given wave vector $k_{z}$, we collect the lowest four eigenfunctions with total angular momentum $|F_{z}|=1/2$, i.e., $\Psi_{1,k_{z},\Uparrow}$, $\Psi_{1,k_{z},\Downarrow}$, $\Psi_{2,k_{z},\Uparrow}$, and $\Psi_{2,k_{z},\Downarrow}$, which span the quasi-degenerate Hilbert subspace in our following perturbation calculations (see appendix \ref{Appendix_a}).

\section{1D hole gas in a strong longitudinal magnetic field}
We now apply a strong longitudinal magnetic field ${\bf B}=(0,0,B)$ to the 1D hole gas. In addition to adding a bare Zeeman term $2\kappa\mu_{B}BJ_{z}$ to the hole Hamiltonian (\ref{Eq_model}), we also need to make the following replacement on the momentum operator ${\bf p}\rightarrow{\bf p+e{\bf A}}$. Here $\kappa=3.41$ is the Luttinger magnetic constant for semiconductor Ge~\cite{PhysRevB.4.3460} and ${\bf A}=(-By/2,Bx/2,0)$ is the vector potential. We can rewrite the hole Hamiltonian as $H=H_{0}+H^{(p)}$, where the zeroth order Hamiltonian $H_{0}$ is given by Eq.~(\ref{Eq_model}), and $H^{(p)}$ is the perturbation Hamiltonian consisting of both the bare Zeeman term and all the orbital terms of the magnetic field (for details see appendix \ref{Appendix_b}).

\begin{figure}
\includegraphics{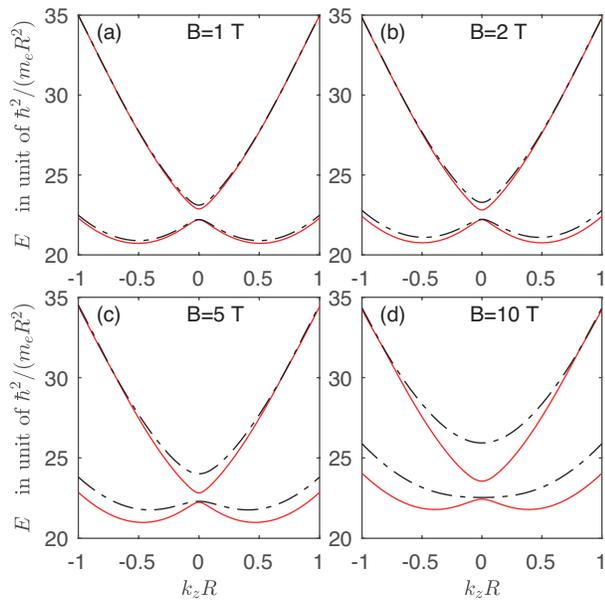}
\caption{\label{Fig_subbands_L}The lowest four subband dispersions of the 1D hole gas under strong longitudinal magnetic fields. The results for $B=1$ T (a), $B=2$ T (b), $B=5$ T (c), and $B=10$ T (d).}
\end{figure}
\begin{figure}
\includegraphics{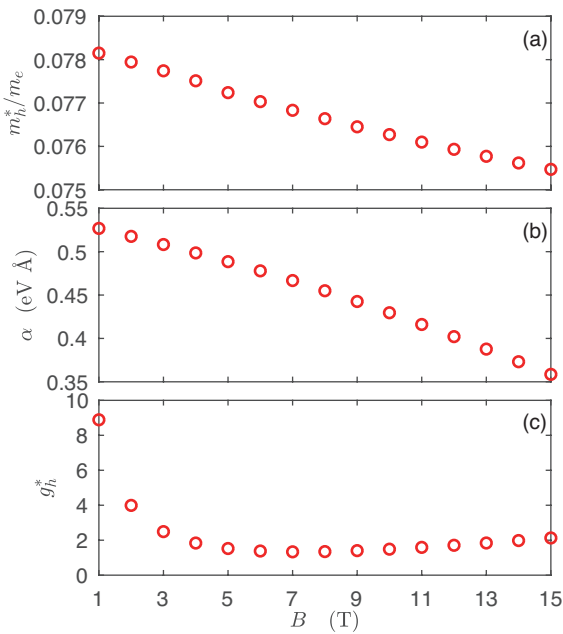}
\caption{\label{Fig_Parameters_L}The magnetic field dependence of the hole effective mass (a), the hole `spin'-orbit coupling (b), and  the effective $g$-factor (c) of the hole `spin'. The magnetic field is applied along the nanowire.}
\end{figure}

The presence of the magnetic field ${\bf B}$ breaks the time-reversal symmetry of the hole Hamiltonian, such that the spin degeneracy in the hole subband dispersions (given in Fig.~\ref{fig_subbands}) is expected to be lifted by this field. In most cases, the hole property is determined by its low-energy subband dispersion, such that here we only focus on the Hilbert subspace spanned by the lowest four subband wave-functions, i.e., $\Psi_{1,k_{z},\Uparrow}$, $\Psi_{1,k_{z},\Downarrow}$, $\Psi_{2,k_{z},\Uparrow}$, and $\Psi_{2,k_{z},\Downarrow}$. We use quasi-degenerate perturbation theory,  i.e., the Hamiltonian $H=H_{0}+H^{(p)}$ is written as a $4\times4$ matrix in this Hilbert subspace (for details see appendix \ref{Appendix_b}), to calculate the splittings in the lowest two subband dispersions. 

The results under various strong longitudinal magnetic fields are shown in Fig.~\ref{Fig_subbands_L}. We also note that quasi-degenerate perturbation calculations in a strong magnetic field do not change the hole $g$-factor at the site $k_{z}=0$. If we label the four subband dispersions in Fig.~\ref{Fig_subbands_L} as $E_{1,2,3,4}(k_{z})$ in sequence from bottom to top, our calculations indicate that $(E_{4}(0)-E_{3}(0))/(\mu_{B}B)=3.13$ (the longitudinal $g$-factor of the second lowest subband~\cite{RL_2021} in Fig.~\ref{fig_subbands}) and $(E_{2}(0)-E_{1}(0))/(\mu_{B}B)=0.14$ (the longitudinal $g$-factor of the lowest subband~\cite{PhysRevB.84.195314,RL_2021} in Fig.~\ref{fig_subbands}) are always satisfied in the magnetic field interval considered here. As we increase the magnetic field, there is an eye visible splitting in the original two-fold degenerate subband dispersions. Now, the low-energy physics of the hole can be represented by the first and the third lowest subband dispersions, i.e., the solid lines given in Fig.~\ref{Fig_subbands_L}. In particular, these two dispersions are well described by the following functional relation
\begin{equation}
\frac{E}{\hbar^{2}/(m_{e}R^{2})}=ak^{2}_{z}R^{2}+b\sigma^{z}k_{z}R+c\sigma^{x}+d,\label{Eq_dispersion_I}
\end{equation}
where $a$, $b$, $c$, and $d$ are dimensionless parameters determined by a second order polynomial curve fitting to the data shown in Fig.~\ref{Fig_subbands_L}, and $\sigma^{x,z}$ are Pauli matrices. We rewrite Eq.~(\ref{Eq_dispersion_I}) in the more intuitional form 
\begin{equation}
E=\frac{\hbar^{2}k^{2}_{z}}{2m^{*}_{h}}+\alpha\sigma^{z}k_{z}+\frac{g^{*}_{h}\mu_{B}B}{2}\sigma^{x}+const.,\label{Eq_dispersion_II}
\end{equation}
where $m^{*}_{h}=m_{e}/(2a)$ can be regarded as the effective hole mass, $\alpha=b\hbar^{2}/(m_{e}R)$ is the strength of the `spin'-orbit coupling, and $g^{*}_{h}=2c\hbar^{2}/(\mu_{B}Bm_{e}R^{2})$ is the effective $g$-factor of the hole `spin'. Note that we have used a strong magnetic field to split off the unwanted spin degree of freedom from the original two-fold degenerate hole subband dispersions, such that the operator $\sigma^{x}$ here does not represent the real hole spin. It is more proper to regard $\sigma^{x}$ as a pseudo spin, this is why we have added single quotes to the word `spin'.

We show the magnetic field dependences of  the effective mass $m^{*}_{h}$, the `spin'-orbit coupling $\alpha$, and the effective $g$-factor $g^{*}_{h}$ of the hole `spin' in Fig.~\ref{Fig_Parameters_L}. Note that the induced effective hole mass $m^{*}_{h}$ in the cylindrical nanowire considered here has the same order of magnitude as that in the planar Ge quantum well~\cite{PhysRevB.103.125201,PhysRevB.4.3460}. The induced spin-orbit coupling given in Eq.~(\ref{Eq_dispersion_II}) is of the linear Rashba type~\cite{Bychkov_1984}, and its magnitude is in the order of several fractions of eV~\AA [see Fig.~\ref{Fig_Parameters_L}(b)]. In the magnetic field interval $1~{\rm T}<B<15~{\rm T}$ considered here,  both $m^{*}_{h}$ and $\alpha$ decrease with the increase of the longitudinal magnetic field. Of course, the variations of $m^{*}_{h}$ and $\alpha$ are small in this interval. While for the effective $g$-factor $g^{*}_{h}$, it only has a small dependence on the magnetic field in the large field region [see Fig.~\ref{Fig_Parameters_L}(c)]. 

Let us discuss the effects of the nanowire radius $R$.  Before the discussion, we emphasize that the plot given in Fig.~\ref{fig_subbands} is independent of $R$~\cite{PhysRevB.84.195314,RL_2021}. First, we consider the impacts of the radius $R$ in the absence of the magnetic field. If we ignore temporarily the spin degeneracy in Fig.~\ref{fig_subbands}, and still model the lowest two subband dispersions as Eqs.~(\ref{Eq_dispersion_I}) and (\ref{Eq_dispersion_II}). The spin-orbit coupling $\alpha=b\hbar^{2}/(m_{e}R)$ is inversely proportional to the nanowire radius $R$, hence it is feasible to increase the absolute strength of the spin-orbit coupling $\alpha$ via reducing $R$. However, the Zeeman splitting $g^{*}_{h}\mu_{B}B\equiv2c\hbar^{2}/(m_{e}R^{2})$ also increases, it follows that the relative strength of the spin-orbit coupling characterized by $m^{*}_{h}\alpha^{2}/(g^{*}_{h}\mu_{B}B\hbar^{2})$ is independent of the nanowire radius $R$. Second, we consider the lifting of the spin degeneracy in Fig.~\ref{fig_subbands} via applying strong magnetic fields. Because we have treated all the magnetic terms (see appendix \ref{Appendix_b}) as perturbations, and the energy unit in Fig.~\ref{fig_subbands} is $\hbar^{2}/(m_{e}R^{2})$, such that the perturbation parameter is proportional to $\mu_{B}B/(\hbar^{2}/(m_{e}R^{2}))$. For smaller nanowire radius, e.g., $R<10$ nm, if we want to achieve the same resolution of the spin splittings as those shown in Fig.~\ref{Fig_subbands_L}, we have to use a series of much larger magnetic fields in order to hold the same order of the perturbation parameter.



\section{1D hole gas in a strong transverse magnetic field}
\begin{figure}
\includegraphics{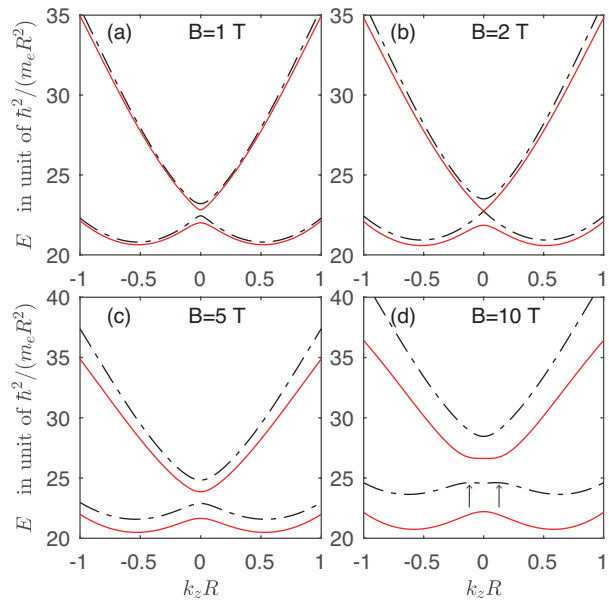}
\caption{\label{Fig_subbands_T}The lowest four subband dispersions of the 1D hole gas under strong transverse magnetic fields. The results for $B=1$ T (a), $B=2$ T (b), $B=5$ T (c), and $B=10$ T (d). The two arrows in (d) mark the two very small anticrossings between the second and the third lowest subband dispersions.}
\end{figure}

\begin{figure}
\includegraphics{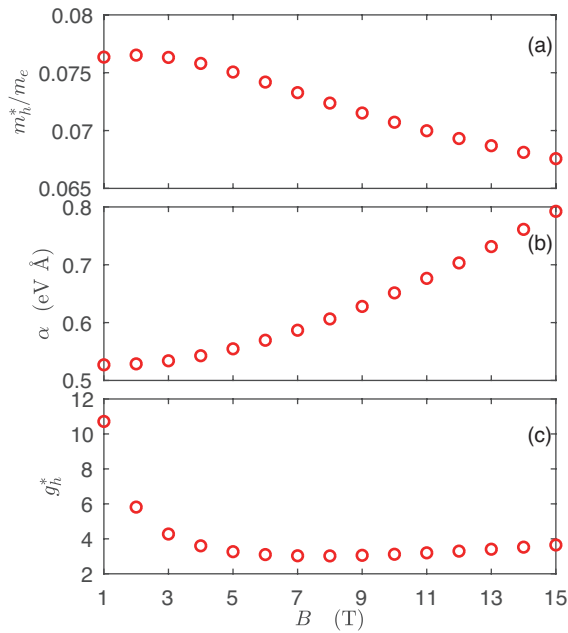}
\caption{\label{Fig_Parameters_T}The magnetic field dependence of the hole effective mass (a), the hole `spin'-orbit coupling (b), and  the  effective $g$-factor (c) of the hole `spin'. The field is applied perpendicular to the nanowire.}
\end{figure}

A strong transverse magnetic field ${\bf B}=(B,0,0)$ is certainly also feasible to split off the unwanted spin degeneracy from the subband dispersions of the 1D hole gas. Now the bare Zeeman term is written as $2\kappa\mu_{B}BJ_{x}$, and  the vector potential can be conveniently chosen as ${\bf A}=(0,0,By)$. By choosing this gauge, the operator $p_{z}$ in the hole Hamiltonian $H$ is still a conserved quantity, and the subband dispersions can still be written $E_{n}(k_{z})$. We rewrite the hole Hamiltonian in perturbative series with respect to the magnetic field $H=H_{0}+H^{(p)}$, where $H^{(p)}$ is the perturbation Hamiltonian consisting of the bare Zeeman term plus all the orbital terms of the magnetic field (for details see appendix \ref{Appendix_c}). Note that the total perturbation Hamiltonian in a transverse field is completely different from that in a longitudinal field.

We use the quasi-degenerate perturbation theory to calculate the lowest four subband dispersions of the 1D hole gas in this strong transverse magnetic field. The obtained results are explicitly shown in Fig.~\ref{Fig_subbands_T}. Because of the large transverse effective hole $g$-factors at $k_{z}R=0$~\cite{PhysRevB.84.195314,RL_2021}, at the magnetic field $B\approx2$ T, the second and the third lowest subband dispersions almost touch with each other at the site $k_{z}R=0$ [see Fig.~\ref{Fig_subbands_T}(b)]. Also, at the site $k_{z}=0$, our calculations indicate that $(E_{4}(0)-E_{2}(0))/(\mu_{B}B)=5.10$ (the transverse $g$-factor of the second lowest subband~\cite{RL_2021} in Fig.~\ref{fig_subbands}) and $(E_{3}(0)-E_{1}(0))/(\mu_{B}B)=5.82$ (the transverse $g$-factor of the lowest subband~\cite{PhysRevB.84.195314,RL_2021} in Fig.~\ref{fig_subbands}) are always satisfied for magnetic fields $B>2$ T. While for magnetic fields $B<2$ T, we have $(E_{4}(0)-E_{3}(0))/(\mu_{B}B)=5.10$ and $(E_{2}(0)-E_{1}(0))/(\mu_{B}B)=5.82$ instead. The above results indicate that there are two very small anticrossings between the second and the third lowest subband dispersions when $B>2$ T, where one anticrossing has a very small negative $k_{z}R$ value and the other anticrossing has a very small positive $k_{z}R$ value [this can be seen clearly when the magnetic field is large enough, e.g., see Fig.~\ref{Fig_subbands_T}(d)]. 

If we ignore the potential consequences of these two very small anticrossings between the second and the third lowest subband dispersions, the first and the third lowest subband dispersions, i.e., the solid lines shown in Fig.~\ref{Fig_subbands_T}, can still be approximately described by Eq.~(\ref{Eq_dispersion_II}). Hence, we still can achieve a strong spin-orbit coupled 1D hole gas in a strong transverse magnetic field. We show the magnetic field dependences of the effective mass $m^{*}_{h}$, the `spin'-orbit coupling $\alpha$, and the effective $g$-factor of the hole `spin' in Fig.~\ref{Fig_Parameters_T}. We note that, different from the longitudinal magnetic field case, with the increase the of the transverse magnetic field, the spin-orbit coupling $\alpha$ increases instead [see Fig.~\ref{Fig_Parameters_T}(b)].

For magnetic fields $B>2$ T, the presence of the two small anticrossings indicates that there exists interplay between the second and the third lowest subband dispersions. However, if the hole energies involved in a special application are far away from these two anticrossings and are close to bottom of the subband dispersions, i.e., close to the minimums of the lowest subband dispersion, it is expected to be a good approximation to neglect the interplay between the second and the third lowest subband dispersions.

Because of the existence of two very small anticrossings between the second and third lowest subband dispersions of the hole gas in a strong transverse magnetic field,  the longitudinal magnetic field may be more proper for achieving the strong `spin'-orbit coupled 1D hole gas.

\section{Discussion and Summary}
In our calculations, we first obtain the exact low-energy subband wave-functions of the model (\ref{Eq_model}), i.e., given by Eq.~(\ref{eq_appendix1}) in appendix~\ref{Appendix_a}, for a series of wave vectors $|k_{z}R|<1$. Then, for each $k_{z}R$ value in this interval, we use quasi-degenerate perturbation theory (the magnetic field is treated perturbatively) to evaluate the spin splittings in the degenerate subband dispersions. While Ref.~\cite{PhysRevB.84.195314} obtained only the low-energy subband wave-functions at the site $k_{z}R=0$, then the Hamiltonian (\ref{Eq_model}) was projected to the Hilbert subspace spanned by these wave-functions at $k_{z}R=0$, in the meantime treating $k_{z}$ in the Hamiltonian (\ref{Eq_model}) as perturbation. Obviously, our results are valid for large $k_{z}R$'s, which may be beyond the validity of Ref.~\cite{PhysRevB.84.195314}. Also, our results at the site $k_{z}R=0$ for both the longitudinal and the transverse field cases agree well with that given in Ref.~\cite{PhysRevB.84.195314}. 

In summary, in the absence of the magnetic field, the Hamiltonian of the hole in a cylindrical nanowire has both the time reversal symmetry and the spin-rotation symmetry, such that there is a spin degeneracy in the induced hole subband dispersions. In this paper, we unambiguously show that a strong `spin'-orbit coupled 1D hole gas is indeed realizable via applying a strong magnetic field to lift the spin degeneracy in the original two-fold degenerate hole subbands. The induced subband dispersion governing the low-energy physics of the hole gas has a very simple form,  which is exactly identical to that of the well studied strong spin-orbit coupled 1D electron gas. We note that the `spin'-orbit coupling obtained here is of the linear Rashba type. We also emphasize that this `spin'-orbit coupling originates from the intrinsic spin-orbit coupling in the Luttinger-Kohn Hamiltonian. It is induced by the subband quantization of the hole gas in a strong magnetic field in the special 1D case. Our study indicates that the strong `spin'-orbit coupled 1D hole gas may have broad applications in comparison with the well-studied 1D  electron gas.

\section*{Acknowledgements}
This work is supported by the National Natural Science Foundation of China Grant No.~11404020, the Project from the Department of Education of Hebei Province Grant No. QN2019057, and the Starting up Foundation from Yanshan University Grant No. BL18043.

\appendix
\begin{widetext}
\section{\label{Appendix_a}Basis states for quasi-degenerate perturbation calculations}
For wave vectors in the interval $|k_{z}R|<1$, the lowest two subbands shown in Fig.~\ref{fig_subbands} are approximately separated from the other higher subbands. The basis states of the quasi-degenerate Hilbert subspace read~\cite{RL_2021}
\begin{equation}
|1\rangle=\left(\begin{array}{c}\Psi_{I,1}(r)e^{-i\varphi}\\\Psi_{I,2}(r)\\\Psi_{I,3}(r)e^{i\varphi}\\\Psi_{I,4}(r)e^{2i\varphi}\end{array}\right),~|2\rangle=\left(\begin{array}{c}\Psi^{*}_{I,4}(r)e^{-2i\varphi}\\\Psi^{*}_{I,3}(r)e^{-i\varphi}\\\Psi^{*}_{I,2}(r)\\\Psi^{*}_{I,1}(r)e^{i\varphi}\end{array}\right),~|3\rangle=\left(\begin{array}{c}\Psi_{II,1}(r)e^{-i\varphi}\\\Psi_{II,2}(r)\\\Psi_{II,3}(r)e^{i\varphi}\\\Psi_{II,4}(r)e^{2i\varphi}\end{array}\right),~|4\rangle=\left(\begin{array}{c}\Psi^{*}_{II,4}(r)e^{-2i\varphi}\\\Psi^{*}_{II,3}(r)e^{-i\varphi}\\\Psi^{*}_{II,2}(r)\\\Psi^{*}_{II,1}(r)e^{i\varphi}\end{array}\right).\label{eq_appendix1}
\end{equation}
In consistence with our preceding notations, here $|1\rangle\equiv\Psi_{1,k_{z},\Uparrow}$, $|2\rangle\equiv\Psi_{1,k_{z},\Downarrow}$, $|3\rangle\equiv\Psi_{2,k_{z},\Uparrow}$, and $ |4\rangle\equiv\Psi_{2,k_{z},\Downarrow}$.

\section{\label{Appendix_b}Longitudinal magnetic field case}
When the magnetic field is applied longitudinally, the hole Hamiltonian can be written in perturbation series 
\begin{equation}
H=H_{0}+H^{(p_{1})}+H^{(p_{2})}+H^{(p_{3})}+H^{(p_{4})}+H^{(p_{5})},
\end{equation}
where $H_{0}$ is the zero-order Hamiltonian given by Eq.~(\ref{Eq_model}), and $H^{(p_{i})}$ ($i=1,\cdots,5$) are the perturbation Hamiltonians. The explicit form of each perturbation term $H^{(p_{i})}$ will be given in the following corresponding subsection. We use quasi-degenerate perturbation theory to calculate the splittings in the original two-fold degenerate subband dispersions. Hence, the total Hamiltonian $H$ is written as a $4\times4$ matrix in the Hilbert subspace spanned by the basis states given in Eq.~(\ref{eq_appendix1}). The zero-order Hamiltonian $H_{0}$ is naturally diagonalized, and the matrix elements of each perturbation term $H^{(p)}$ are given in the following.
 
\subsection{Perturbation term I: $H^{(p_{1})}=2\kappa\mu_{B}BJ_{z}$}
The matrix elements of the perturbation term $H^{(p_{1})}$ read
\begin{eqnarray}
H^{(p_{1})}_{11}&=&4\pi\kappa\mu_{B}B\int^{R}_{0}drr\Big(\frac{3}{2}|\Psi_{I,1}(r)|^{2}+\frac{1}{2}|\Psi_{I,2}(r)|^{2}-\frac{1}{2}|\Psi_{I,3}(r)|^{2}-\frac{3}{2}|\Psi_{I,4}(r)|^{2}\Big),\nonumber\\
H^{(p_{1})}_{12}&=&0,\nonumber\\
H^{(p_{1})}_{13}&=&4\pi\kappa\mu_{B}B\int^{R}_{0}dr\,r\Big(\frac{3}{2}\Psi^{*}_{I,1}(r)\Psi_{II,1}(r)+\frac{1}{2}\Psi^{*}_{I,2}(r)\Psi_{II,2}(r)-\frac{1}{2}\Psi^{*}_{I,3}(r)\Psi_{II,3}(r)-\frac{3}{2}\Psi^{*}_{I,4}(r)\Psi_{II,4}(r)\Big),\nonumber\\
H^{(p_{1})}_{14}&=&0,\nonumber\\
H^{(p_{1})}_{22}&=&-H^{(p_{1})}_{11},\nonumber\\
H^{(p_{1})}_{23}&=&0,\nonumber\\
H^{(p_{1})}_{24}&=&-\left(H^{(p_{1})}_{13}\right)^{*},\nonumber\\
H^{(p_{1})}_{33}&=&4\pi\kappa\mu_{B}B\int^{R}_{0}drr\Big(\frac{3}{2}|\Psi_{II,1}(r)|^{2}+\frac{1}{2}|\Psi_{II,2}(r)|^{2}-\frac{1}{2}|\Psi_{II,3}(r)|^{2}-\frac{3}{2}|\Psi_{II,4}(r)|^{2}\Big),\nonumber\\
H^{(p_{1})}_{34}&=&0,\nonumber\\
H^{(p_{1})}_{44}&=&-H^{(p_{1})}_{33}.
\end{eqnarray}

\subsection{Perturbation term II: $H^{(p_{2})}$}
In this subsection, the perturbation term reads
\begin{equation}
H^{(p_{2})}=-i\mu_{B}B\partial_{\varphi}\left(\begin{array}{cccc}\gamma_{1}+\gamma_{s}&0&0&0\\0&\gamma_{1}-\gamma_{s}&0&0\\0&0&\gamma_{1}-\gamma_{s}&0\\0&0&0&\gamma_{1}+\gamma_{s}\end{array}\right).
\end{equation}
The matrix elements of the perturbation term read
\begin{eqnarray}
H^{(p_{2})}_{11}&=&2\pi\mu_{B}B\int^{R}_{0}drr\Big(-(\gamma_{1}+\gamma_{s})|\Psi_{I,1}(r)|^{2}+(\gamma_{1}-\gamma_{s})|\Psi_{I,3}(r)|^{2}+2(\gamma_{1}+\gamma_{s})|\Psi_{I,4}(r)|^{2}\Big),\nonumber\\
H^{(p_{2})}_{12}&=&0,\nonumber\\
H^{(p_{2})}_{13}&=&2\pi\mu_{B}B\int^{R}_{0}dr\,r\Big(-(\gamma_{1}+\gamma_{s})\Psi^{*}_{I,1}(r)\Psi_{II,1}(r)+(\gamma_{1}-\gamma_{s})\Psi^{*}_{I,3}(r)\Psi_{II,3}(r)+2(\gamma_{1}+\gamma_{s})\Psi^{*}_{I,4}(r)\Psi_{II,4}(r)\Big),\nonumber\\
H^{(p_{2})}_{14}&=&0,\nonumber\\
H^{(p_{2})}_{22}&=&-H^{(p_{2})}_{11},\nonumber\\
H^{(p_{2})}_{23}&=&0,\nonumber\\
H^{(p_{2})}_{24}&=&-\left(H^{(p_{2})}_{13}\right)^{*},\nonumber\\
H^{(p_{2})}_{33}&=&2\pi\mu_{B}B\int^{R}_{0}drr\Big(-(\gamma_{1}+\gamma_{s})|\Psi_{II,1}(r)|^{2}+(\gamma_{1}-\gamma_{s})|\Psi_{II,3}(r)|^{2}+2(\gamma_{1}+\gamma_{s})|\Psi_{II,4}(r)|^{2}\Big),\nonumber\\
H^{(p_{2})}_{34}&=&0,\nonumber\\
H^{(p_{2})}_{44}&=&-H^{(p_{2})}_{33}.
\end{eqnarray}

\subsection{Perturbation term III: $H^{(p_{3})}$}
In this subsection, the perturbation term reads
\begin{equation}
H^{(p_{3})}=-\sqrt{3}\gamma_{s}\mu_{B}B\left(\begin{array}{cccc}0&-ik_{z}re^{-i\varphi}&e^{-2i\varphi}(-r\partial_{r}+i\partial_{\varphi})&0\\ik_{z}re^{i\varphi}&0&0&e^{-2i\varphi}(-r\partial_{r}+i\partial_{\varphi})\\e^{2i\varphi}(r\partial_{r}+i\partial_{\varphi})&0&0&ik_{z}re^{-i\varphi}\\0&e^{2i\varphi}(r\partial_{r}+i\partial_{\varphi})&-ik_{z}re^{i\varphi}&0\end{array}\right).
\end{equation}
The matrix elements of the perturbation term read
\begin{eqnarray}
H^{(p_{3})}_{11}&=&2\pi\sqrt{3}\gamma_{s}\mu_{B}B\int^{R}_{0}drr\Big(ik_{z}r\Psi^{*}_{I,1}(r)\Psi_{I,2}(r)-ik_{z}r\Psi^{*}_{I,2}(r)\Psi_{I,1}(r)-ik_{z}r\Psi^{*}_{I,3}(r)\Psi_{I,4}(r)\nonumber\\
&&+ik_{z}r\Psi^{*}_{I,4}(r)\Psi_{I,3}(r)+\Psi^{*}_{I,1}(r)(r\partial_{r}+1)\Psi_{I,3}(r)+\Psi_{I,1}(r)(r\partial_{r}+1)\Psi^{*}_{I,3}(r)\nonumber\\
&&+\Psi^{*}_{I,2}(r)(r\partial_{r}+2)\Psi_{I,4}(r)+\Psi_{I,2}(r)(r\partial_{r}+2)\Psi^{*}_{I,4}(r)\Big),\nonumber\\
H^{(p_{3})}_{12}&=&0,\nonumber\\
H^{(p_{3})}_{13}&=&2\pi\sqrt{3}\gamma_{s}\mu_{B}B\int^{R}_{0}dr\,r\Big(ik_{z}r\Psi^{*}_{I,1}(r)\Psi_{II,2}(r)-ik_{z}r\Psi^{*}_{I,2}(r)\Psi_{II,1}(r)-ik_{z}r\Psi^{*}_{I,3}(r)\Psi_{II,4}(r)\nonumber\\
&&+ik_{z}r\Psi^{*}_{I,4}(r)\Psi_{II,3}(r)+\Psi^{*}_{I,1}(r)(r\partial_{r}+1)\Psi_{II,3}(r)+\Psi_{II,1}(r)(r\partial_{r}+1)\Psi^{*}_{I,3}(r)\nonumber\\
&&+\Psi^{*}_{I,2}(r)(r\partial_{r}+2)\Psi_{II,4}(r)+\Psi_{II,2}(r)(r\partial_{r}+2)\Psi^{*}_{I,4}(r)\Big),\nonumber\\
H^{(p_{3})}_{14}&=&0,\nonumber\\
H^{(p_{3})}_{22}&=&2\pi\sqrt{3}\gamma_{s}\mu_{B}B\int^{R}_{0}drr\Big(-ik_{z}r\Psi_{I,2}(r)\Psi^{*}_{I,1}(r)+ik_{z}r\Psi_{I,1}(r)\Psi^{*}_{I,2}(r)+ik_{z}r\Psi_{I,4}(r)\Psi^{*}_{I,3}(r)\nonumber\\
&&-ik_{z}r\Psi_{I,3}(r)\Psi^{*}_{I,4}(r)-\Psi_{I,1}(r)(r\partial_{r}+1)\Psi^{*}_{I,3}(r)-\Psi^{*}_{I,1}(r)(r\partial_{r}+1)\Psi_{I,3}(r)\nonumber\\
&&-\Psi_{I,2}(r)(r\partial_{r}+2)\Psi^{*}_{I,4}(r)-\Psi^{*}_{I,2}(r)(r\partial_{r}+2)\Psi_{I,4}(r)\Big),\nonumber\\
H^{(p_{3})}_{23}&=&0,\nonumber\\
H^{(p_{3})}_{24}&=&-2\pi\sqrt{3}\gamma_{s}\mu_{B}B\int^{R}_{0}drr\Big(-ik_{z}r\Psi_{I,1}(r)\Psi^{*}_{II,2}(r)+ik_{z}r\Psi_{I,2}(r)\Psi^{*}_{II,1}(r)+ik_{z}r\Psi_{I,3}(r)\Psi^{*}_{II,4}(r)\nonumber\\
&&-ik_{z}r\Psi_{I,4}(r)\Psi^{*}_{II,3}(r)+\Psi^{*}_{II,2}(r)(r\partial_{r}+2)\Psi_{I,4}(r)+\Psi^{*}_{II,1}(r)(r\partial_{r}+1)\Psi_{I,3}(r)\nonumber\\
&&+\Psi_{I,2}(r)(r\partial_{r}+2)\Psi^{*}_{II,4}(r)+\Psi_{I,1}(r)(r\partial_{r}+1)\Psi^{*}_{II,3}(r)\Big),\nonumber\\
H^{(p_{3})}_{33}&=&2\pi\sqrt{3}\gamma_{s}\mu_{B}B\int^{R}_{0}drr\Big(ik_{z}r\Psi^{*}_{II,1}(r)\Psi_{II,2}(r)-ik_{z}r\Psi^{*}_{II,2}(r)\Psi_{II,1}(r)-ik_{z}r\Psi^{*}_{II,3}(r)\Psi_{II,4}(r)\nonumber\\
&&+ik_{z}r\Psi^{*}_{II,4}(r)\Psi_{II,3}(r)+\Psi^{*}_{II,1}(r)(r\partial_{r}+1)\Psi_{II,3}(r)+\Psi_{II,1}(r)(r\partial_{r}+1)\Psi^{*}_{II,3}(r)\nonumber\\
&&+\Psi^{*}_{II,2}(r)(r\partial_{r}+2)\Psi_{II,4}(r)+\Psi_{II,2}(r)(r\partial_{r}+2)\Psi^{*}_{II,4}(r)\Big),\nonumber\\
H^{(p_{3})}_{34}&=&0,\nonumber\\
H^{(p_{3})}_{44}&=&2\pi\sqrt{3}\gamma_{s}\mu_{B}B\int^{R}_{0}drr\Big(-ik_{z}r\Psi_{II,2}(r)\Psi^{*}_{II,1}(r)+ik_{z}r\Psi_{II,1}(r)\Psi^{*}_{II,2}(r)+ik_{z}r\Psi_{II,4}(r)\Psi^{*}_{II,3}(r)\nonumber\\
&&-ik_{z}r\Psi_{II,3}(r)\Psi^{*}_{II,4}(r)-\Psi_{II,1}(r)(r\partial_{r}+1)\Psi^{*}_{II,3}(r)-\Psi^{*}_{II,1}(r)(r\partial_{r}+1)\Psi_{II,3}(r)\nonumber\\
&&-\Psi_{II,2}(r)(r\partial_{r}+2)\Psi^{*}_{II,4}(r)-\Psi^{*}_{II,2}(r)(r\partial_{r}+2)\Psi_{II,4}(r)\Big).
\end{eqnarray}

\subsection{Perturbation term IV: $H^{(p_{4})}$}
In this subsection, the perturbation term reads
\begin{equation}
H^{(p_{4})}=\left(\gamma_{1}+\frac{5}{2}\gamma_{s}\right)\frac{e^{2}B^{2}r^{2}}{8m_{e}}.
\end{equation}
The matrix elements of the perturbation term read
\begin{eqnarray}
H^{(p_{4})}_{11}&=&2\pi(\gamma_{1}+\frac{5}{2}\gamma_{s})\frac{e^{2}B^{2}}{8m_{e}}\int^{R}_{0}drr^{3}\Big(|\Psi_{I,1}(r)|^{2}+|\Psi_{I,2}(r)|^{2}+|\Psi_{I,3}(r)|^{2}+|\Psi_{I,4}(r)|^{2}\Big),\nonumber\\
H^{(p_{4})}_{12}&=&0,\nonumber\\
H^{(p_{4})}_{13}&=&2\pi\left(\gamma_{1}+\frac{5}{2}\gamma_{s}\right)\frac{e^{2}B^{2}}{8m_{e}}\int^{R}_{0}drr^{3}\Big(\Psi^{*}_{I,1}(r)\Psi_{II,1}(r)+\Psi^{*}_{I,2}(r)\Psi_{II,2}(r)+\Psi^{*}_{I,3}(r)\Psi_{II,3}(r)+\Psi^{*}_{I,4}(r)\Psi_{II,4}(r)\Big),\nonumber\\
H^{(p_{4})}_{14}&=&0,\nonumber\\
H^{(p_{4})}_{22}&=&H^{(p_{4})}_{11},\nonumber\\
H^{(p_{4})}_{23}&=&0,\nonumber\\
H^{(p_{4})}_{24}&=&\left(H^{(p_{4})}_{13}\right)^{*},\nonumber\\
H^{(p_{4})}_{33}&=&2\pi(\gamma_{1}+\frac{5}{2}\gamma_{s})\frac{e^{2}B^{2}}{8m_{e}}\int^{R}_{0}drr^{3}\Big(|\Psi_{II,1}(r)|^{2}+|\Psi_{II,2}(r)|^{2}+|\Psi_{II,3}(r)|^{2}+|\Psi_{II,4}(r)|^{2}\Big),\nonumber\\
H^{(p_{4})}_{34}&=&0,\nonumber\\
H^{(p_{4})}_{44}&=&H^{(p_{4})}_{33}.
\end{eqnarray}

\subsection{Perturbation term V: $H^{(p_{5})}$}
In this subsection, the perturbation term reads
\begin{equation}
H^{(p_{5})}=-\frac{\gamma_{s}e^{2}B^{2}r^{2}}{4m_{e}}\left(\begin{array}{cccc}\frac{3}{4}&0&-\frac{\sqrt{3}}{2}e^{-2i\varphi}&0\\0&\frac{7}{4}&0&-\frac{\sqrt{3}}{2}e^{-2i\varphi}\\-\frac{\sqrt{3}}{2}e^{2i\varphi}&0&\frac{7}{4}&0\\0&-\frac{\sqrt{3}}{2}e^{2i\varphi}&0&\frac{3}{4}\end{array}\right).
\end{equation}
The matrix elements of the perturbation term read
\begin{eqnarray}
H^{(p_{5})}_{11}&=&-\frac{2\pi\gamma_{s}e^{2}B^{2}}{4m_{e}}\int^{R}_{0}drr^{3}\Big(-\frac{\sqrt{3}}{2}\Psi^{*}_{I,1}(r)\Psi_{I,3}(r)-\frac{\sqrt{3}}{2}\Psi^{*}_{I,3}(r)\Psi_{I,1}(r)-\frac{\sqrt{3}}{2}\Psi^{*}_{I,2}(r)\Psi_{I,4}(r)-\frac{\sqrt{3}}{2}\Psi^{*}_{I,4}(r)\Psi_{I,2}(r)\nonumber\\
&&+\frac{3}{4}|\Psi_{I,1}(r)|^{2}+\frac{7}{4}|\Psi_{I,2}(r)|^{2}+\frac{7}{4}|\Psi_{I,3}(r)|^{2}+\frac{3}{4}|\Psi_{I,4}(r)|^{2}\Big),\nonumber\\
H^{(p_{5})}_{12}&=&0,\nonumber\\
H^{(p_{5})}_{13}&=&-\frac{2\pi\gamma_{s}e^{2}B^{2}}{4m_{e}}\int^{R}_{0}drr^{3}\Big(-\frac{\sqrt{3}}{2}\Psi^{*}_{I,1}(r)\Psi_{II,3}(r)-\frac{\sqrt{3}}{2}\Psi^{*}_{I,3}(r)\Psi_{II,1}(r)-\frac{\sqrt{3}}{2}\Psi^{*}_{I,2}(r)\Psi_{II,4}(r)\nonumber\\
&&-\frac{\sqrt{3}}{2}\Psi^{*}_{I,4}(r)\Psi_{II,2}(r)+\frac{3}{4}\Psi^{*}_{I,1}(r)\Psi_{II,1}(r)+\frac{7}{4}\Psi^{*}_{I,2}(r)\Psi_{II,2}(r)+\frac{7}{4}\Psi^{*}_{I,3}(r)\Psi_{II,3}(r)+\frac{3}{4}\Psi^{*}_{I,4}(r)\Psi_{II,4}(r)\Big),\nonumber\\
H^{(p_{5})}_{14}&=&0,\nonumber\\
H^{(p_{5})}_{22}&=&H^{(p_{5})}_{11},\nonumber\\
H^{(p_{5})}_{23}&=&0,\nonumber\\
H^{(p_{5})}_{24}&=&\left(H^{(p_{5})}_{13}\right)^{*},\nonumber\\
H^{(p_{5})}_{33}&=&-\frac{2\pi\gamma_{s}e^{2}B^{2}}{4m_{e}}\int^{R}_{0}drr^{3}\Big(-\frac{\sqrt{3}}{2}\Psi^{*}_{II,1}(r)\Psi_{II,3}(r)-\frac{\sqrt{3}}{2}\Psi^{*}_{II,3}(r)\Psi_{II,1}(r)-\frac{\sqrt{3}}{2}\Psi^{*}_{II,2}(r)\Psi_{II,4}(r)\nonumber\\
&&-\frac{\sqrt{3}}{2}\Psi^{*}_{II,4}(r)\Psi_{II,2}(r)+\frac{3}{4}|\Psi_{II,1}(r)|^{2}+\frac{7}{4}|\Psi_{II,2}(r)|^{2}+\frac{7}{4}|\Psi_{II,3}(r)|^{2}+\frac{3}{4}|\Psi_{II,4}(r)|^{2}\Big),\nonumber\\
H^{(p_{5})}_{34}&=&0,\nonumber\\
H^{(p_{5})}_{44}&=&H^{(p_{5})}_{33}.
\end{eqnarray}

\section{\label{Appendix_c}Transverse magnetic field case}
When the magnetic field is applied transversely, the hole Hamiltonian can also be written in perturbation series 
\begin{equation}
H=H_{0}+H^{(p_{1})}+H^{(p_{2})}+H^{(p_{3})}+H^{(p_{4})},
\end{equation}
where $H_{0}$ is the zero-order diagonal Hamiltonian, and $H^{(p_{i})}$ ($i=1,\cdots,4$) are the perturbation Hamiltonians. The explicit form of each perturbation term $H^{(p_{i})}$ is given in the following corresponding subsection.

\subsection{Perturbation term I: $H^{(p_{1})}=2\kappa\mu_{B}BJ_{x}$}
The matrix elements of the perturbation term read
\begin{eqnarray}
H^{(p_{1})}_{11}&=&0,\nonumber\\
H^{(p_{1})}_{12}&=&4\pi\kappa\mu_{B}B\int^{R}_{0}drr\Big(\sqrt{3}\Psi^{*}_{I,1}(r)\Psi^{*}_{I,3}(r)+\Psi^{*}_{I,2}(r)\Psi^{*}_{I,2}(r)\Big),\nonumber\\
H^{(p_{1})}_{13}&=&0,\nonumber\\
H^{(p_{1})}_{14}&=&4\pi\kappa\mu_{B}B\int^{R}_{0}drr\Big(\frac{\sqrt{3}}{2}\Psi^{*}_{I,1}(r)\Psi^{*}_{II,3}(r)+\frac{\sqrt{3}}{2}\Psi^{*}_{I,3}(r)\Psi^{*}_{II,1}(r)+\Psi^{*}_{I,2}(r)\Psi^{*}_{II,2}(r)\Big),\nonumber\\
H^{(p_{1})}_{22}&=&0,\nonumber\\
H^{(p_{1})}_{23}&=&\left(H^{(p_{1})}_{14}\right)^{*},\nonumber\\
H^{(p_{1})}_{24}&=&0,\nonumber\\
H^{(p_{1})}_{33}&=&0,\nonumber\\
H^{(p_{1})}_{34}&=&4\pi\kappa\mu_{B}B\int^{R}_{0}drr\Big(\sqrt{3}\Psi^{*}_{II,1}(r)\Psi^{*}_{II,3}(r)+\Psi^{*}_{II,2}(r)\Psi^{*}_{II,2}(r)\Big),\nonumber\\
H^{(p_{1})}_{44}&=&0.
\end{eqnarray}

\subsection{Perturbation term II: $H^{(p_{2})}$}
In this subsection, the perturbation term reads
\begin{equation}
{\footnotesize H^{(p_{2})}=2\sqrt{3}\gamma_{s}\mu_{B}B\left(\begin{array}{cccc}0&e^{-i\varphi}\sin\varphi(ir\partial_{r}+\partial_{\varphi})+\frac{1}{2}&0&0\\e^{i\varphi}\sin\varphi(ir\partial_{r}-\partial_{\varphi})-\frac{1}{2}&0&0&0\\0&0&0&e^{-i\varphi}\sin\varphi(-ir\partial_{r}-\partial_{\varphi})-\frac{1}{2}\\0&0&e^{i\varphi}\sin\varphi(-ir\partial_{r}+\partial_{\varphi})+\frac{1}{2}&0\end{array}\right).}
\end{equation}
The matrix elements of the perturbation term read
\begin{eqnarray}
H^{(p_{2})}_{11}&=&0,\nonumber\\
H^{(p_{2})}_{12}&=&4\pi\sqrt{3}\gamma_{s}\mu_{B}B\int^{R}_{0}drr\Big(\Psi^{*}_{I,1}(r)(r\partial_{r}+1)\Psi^{*}_{I,3}(r)+\Psi^{*}_{I,2}(r)(r\partial_{r}+2)\Psi^{*}_{I,4}(r)-\Psi^{*}_{I,1}(r)\Psi^{*}_{I,3}(r)\Big),\nonumber\\
H^{(p_{2})}_{13}&=&0,\nonumber\\
H^{(p_{2})}_{14}&=&2\pi\sqrt{3}\gamma_{s}\mu_{B}B\int^{R}_{0}drr\Big(\Psi^{*}_{I,1}(r)(r\partial_{r}+1)\Psi^{*}_{II,3}(r)+\Psi^{*}_{II,1}(r)(r\partial_{r}+1)\Psi^{*}_{I,3}(r)+\Psi^{*}_{I,2}(r)(r\partial_{r}+2)\Psi^{*}_{II,4}(r)\nonumber\\
&&+\Psi^{*}_{II,2}(r)(r\partial_{r}+2)\Psi^{*}_{I,4}(r)-\Psi^{*}_{I,1}(r)\Psi^{*}_{II,3}(r)-\Psi^{*}_{II,1}(r)\Psi^{*}_{I,3}(r)\Big),\nonumber\\
H^{(p_{2})}_{22}&=&0,\nonumber\\
H^{(p_{2})}_{23}&=&\left(H^{(p_{2})}_{14}\right)^{*},\nonumber\\
H^{(p_{2})}_{24}&=&0,\nonumber\\
H^{(p_{2})}_{33}&=&0,\nonumber\\
H^{(p_{2})}_{34}&=&4\pi\sqrt{3}\gamma_{s}\mu_{B}B\int^{R}_{0}drr\Big(\Psi^{*}_{II,1}(r)(r\partial_{r}+1)\Psi^{*}_{II,3}(r)+\Psi^{*}_{II,2}(r)(r\partial_{r}+2)\Psi^{*}_{II,4}(r)-\Psi^{*}_{II,1}(r)\Psi^{*}_{II,3}(r)\Big),\nonumber\\
H^{(p_{2})}_{44}&=&0.
\end{eqnarray}

\subsection{Perturbation term III: $H^{(p_{3})}$}
In this subsection, the perturbation term reads
\begin{eqnarray}
H^{(p_{3})}&=&\frac{e^{2}B^{2}r^{2}\sin^{2}\varphi}{2m_{e}}\left(\begin{array}{cccc}\gamma_{1}-2\gamma_{s}&0&0&0\\0&\gamma_{1}+2\gamma_{s}&0&0\\0&0&\gamma_{1}+2\gamma_{s}&0\\0&0&0&\gamma_{1}-2\gamma_{s}\end{array}\right).
\end{eqnarray}
The matrix elements of the perturbation term read
\begin{eqnarray}
H^{(p_{3})}_{11}&=&\frac{\pi\,e^{2}B^{2}}{2m_{e}}\int^{R}_{0}drr^{3}\Big((\gamma_{1}-2\gamma_{s})|\Psi_{I,1}(r)|^{2}+(\gamma_{1}+2\gamma_{s})|\Psi_{I,2}(r)|^{2}+(\gamma_{1}+2\gamma_{s})|\Psi_{I,3}(r)|^{2}+(\gamma_{1}-2\gamma_{s})|\Psi_{I,4}(r)|^{2}\Big),\nonumber\\
H^{(p_{3})}_{12}&=&0,\nonumber\\
H^{(p_{3})}_{13}&=&\frac{\pi\,e^{2}B^{2}}{2m_{e}}\int^{R}_{0}drr^{3}\Big((\gamma_{1}-2\gamma_{s})\Psi^{*}_{I,1}(r)\Psi_{II,1}(r)+(\gamma_{1}+2\gamma_{s})\Psi^{*}_{I,2}(r)\Psi_{II,2}(r)\nonumber\\
&&+(\gamma_{1}+2\gamma_{s})\Psi^{*}_{I,3}(r)\Psi_{II,3}(r)+(\gamma_{1}-2\gamma_{s})\Psi^{*}_{I,4}(r)\Psi_{II,4}(r)\Big),\nonumber\\
H^{(p_{3})}_{14}&=&0,\nonumber\\
H^{(p_{3})}_{22}&=&H^{(p_{3})}_{11},\nonumber\\
H^{(p_{3})}_{23}&=&0,\nonumber\\
H^{(p_{3})}_{24}&=&\left(H^{(p_{3})}_{13}\right)^{*},\nonumber\\
H^{(p_{3})}_{33}&=&\frac{\pi\,e^{2}B^{2}}{2m_{e}}\int^{R}_{0}drr^{3}\Big((\gamma_{1}-2\gamma_{s})|\Psi_{II,1}(r)|^{2}+(\gamma_{1}+2\gamma_{s})|\Psi_{II,2}(r)|^{2}\nonumber\\
&&+(\gamma_{1}+2\gamma_{s})|\Psi_{II,3}(r)|^{2}+(\gamma_{1}-2\gamma_{s})|\Psi_{II,4}(r)|^{2}\Big),\nonumber\\
H^{(p_{3})}_{34}&=&0,\nonumber\\
H^{(p_{3})}_{44}&=&H^{(p_{3})}_{33}.
\end{eqnarray}

\subsection{Perturbation term IV: $H^{(p_{4})}$}
In this subsection, the perturbation term reads
\begin{eqnarray}
H^{(p_{4})}&=&2\mu_{B}Bk_{z}r\sin\varphi\left(\begin{array}{cccc}\gamma_{1}-2\gamma_{s}&0&0&0\\0&\gamma_{1}+2\gamma_{s}&0&0\\0&0&\gamma_{1}+2\gamma_{s}&0\\0&0&0&\gamma_{1}-2\gamma_{s}\end{array}\right).
\end{eqnarray}
The matrix elements of the perturbation term read
\begin{eqnarray}
H^{(p_{4})}_{11}&=&0,\nonumber\\
H^{(p_{4})}_{12}&=&-4i\pi\mu_{B}Bk_{z}\int^{R}_{0}drr^{2}\Big((\gamma_{1}+2\gamma_{s})\Psi^{*}_{I,2}(r)\Psi^{*}_{I,3}(r)+(\gamma_{1}-2\gamma_{s})\Psi^{*}_{I,1}(r)\Psi^{*}_{I,4}(r)\Big),\nonumber\\
H^{(p_{4})}_{13}&=&0,\nonumber\\
H^{(p_{4})}_{14}&=&-2i\pi\mu_{B}Bk_{z}\int^{R}_{0}drr^{2}\Big((\gamma_{1}-2\gamma_{s})\Psi^{*}_{I,1}(r)\Psi^{*}_{II,4}(r)+(\gamma_{1}+2\gamma_{s})\Psi^{*}_{I,2}(r)\Psi^{*}_{II,3}(r)\nonumber\\
&&+(\gamma_{1}+2\gamma_{s})\Psi^{*}_{I,3}(r)\Psi^{*}_{II,2}(r)+(\gamma_{1}-2\gamma_{s})\Psi^{*}_{I,4}(r)\Psi^{*}_{II,1}(r)\Big),\nonumber\\
H^{(p_{4})}_{22}&=&0,\nonumber\\
H^{(p_{4})}_{23}&=&\left(H^{(p_{4})}_{14}\right)^{*},\nonumber\\
H^{(p_{4})}_{24}&=&0,\nonumber\\
H^{(p_{4})}_{33}&=&0,\nonumber\\
H^{(p_{4})}_{34}&=&-4i\pi\mu_{B}Bk_{z}\int^{R}_{0}drr^{2}\Big((\gamma_{1}+2\gamma_{s})\Psi^{*}_{II,2}(r)\Psi^{*}_{II,3}(r)+(\gamma_{1}-2\gamma_{s})\Psi^{*}_{II,1}(r)\Psi^{*}_{II,4}(r)\Big),\nonumber\\
H^{(p_{4})}_{44}&=&0.
\end{eqnarray}
\end{widetext}
\bibliography{Ref_Hole_spin}
\end{document}